\documentclass[%
 reprint,
 amsmath,amssymb,
 aps,
]{revtex4-1}

\usepackage{graphicx}% Include figure files
\usepackage{dcolumn}% Align table columns on decimal point
\usepackage{bm}% bold math
\usepackage{hyperref}% add hypertext capabilities
\usepackage[caption=false]{subfig}
\usepackage{siunitx}
\usepackage{amsthm}
\usepackage{physics}
\usepackage{mathtools}
\usepackage{xcolor}

\DeclareMathOperator*{\argmax}{arg\,max}
\DeclareMathOperator{\variance}{Var}
\DeclarePairedDelimiter{\ceil}{\lceil}{\rceil}
\DeclarePairedDelimiter{\floor}{\lfloor}{\rfloor}

\begin{document}

\preprint{APS/123-QED}

\title{Contrast resolution of few-photon detectors}% Force line breaks with \\
%\thanks{A footnote to the article title}%

\author{Mattias J\"{o}nsson}
\email{matjon4@kth.se}
\affiliation{%
 Department of Physics, KTH Royal Institute of Technology\\
 AlbaNova University Center, SE 106 91 Stockholm, Sweden
}%

% \altaffiliation[Also at ]{Physics Department, XYZ University.}
\author{Gunnar Bj\"{o}rk}%
\email{gbjork@kth.se}
\affiliation{%
 Department of Applied Physics, KTH Royal Institute of Technology\\
 AlbaNova University Center, SE 106 91 Stockholm, Sweden
}%

%\collaboration{MUSO Collaboration}%\noaffiliation

%\author{Charlie Author}
% \homepage{http://www.Second.institution.edu/~Charlie.Author}
%\affiliation{
% Second institution and/or address\\
% This line break forced% with \\
%}%
%\affiliation{
% Third institution, the second for Charlie Author
%}%
%\author{Delta Author}
%\affiliation{%
% Authors' institution and/or address\\
% This line break forced with \textbackslash\textbackslash
%}%

%\collaboration{CLEO Collaboration}%\noaffiliation

\date{\today}% It is always \today, today,
             %  but any date may be explicitly specified

\begin{abstract}
We analyse the capability of distinguishing between different intensities in a monochromatic, pixellated image acquisition system at low light intensities. In practice, the latter means that each pixel detects a countable number of photons per acquired image frame. Primarily we compare systems based on pixels of the click-detection type and photon-number resolving (PNR) type of detectors, but our model can seamlessly interpolate between the two. We also discuss the probability of errors in assigning the correct intensity (or ``gray level''), and finally we discuss how the estimated levels, that need to be based on threshold levels due to the stochastic nature of the detected photon number, should be assigned. Overall, we find that PNR detector-based system offer advantages over click-detector-based systems even under rather non-ideal conditions.
\end{abstract}

\maketitle

%\tableofcontents

\section{\label{sec:introduction}INTRODUCTION}
In many fields of science, medicine, and technology image acquisition plays an important role and imaging constitutes a large and wide research field. A common imaging problem to consider is the reconstruction of a gray-scale image from a, possibly repeated, measurement with an array of photon detector. In an ordinary digital camera this is done after RGB color separation of the incoming image. Hence, even for color imaging the problem boils down to resolution of a narrow wavelength-band ``gray-scales", which necessitates the resolution of contrast between the different levels \cite{Alsleem2012QualityImages}.

The imaging problem becomes more difficult when the illumination is constrained to the few photon level, which is the case for a number of application such as LIDAR \cite{Zhou2015, Shin2016Photon-efficientCamera}, biological spectroscopy \cite{Niwa2017Few-photonSpectrometry}, image-scanning microscopy \cite{ButtafavaSPAD-basedMicroscopy} and low-dose x-ray \cite{Zhu2018Few-photonImaging}. In these cases the use of single-photon detectors such as single-photon avalanche photo-detector (SPADs), superconducting nanowire single-photon detectors (SNSPDs) or transition-edge sensors (TES) are required to resolve the faint signals. Building an imaging device with sufficiently many pixels to generate a high resolution image is still an area of research \cite{Guerrieri2010Two-dimensionalCounting, Bronzi2014100Ranging, Allman2015AReadout, Gottardi2016DevelopmentATHENA, gasparini2018, Wollman2019KilopixelDetectors}.

In recent development it has been shown that some information about the incident photon numbers can be gained from avalanche photo-detectors (APDs) and SNSPDs by analysing the output pulses \cite{Kardyna2008AnDetector, Nicolich2019UniversalDetectors, Zhu2019ResolvingDetector}. Alternatively, multiplexed structures \cite{Divochiy2008, Marsili2009PhysicsNanowires, Dauler2009Photon-number-resolutionDetectors} or inherent photon-number-resolving (PNR) detectors \cite{Lita2008, Fukuda2011, Konno2020DevelopmentArray, Young2020DesignElements} can be used to gain photon-number resolution. Utilizing that these detectors exhibit some PNR capabilities allows for more efficient information extraction of the input and this could improve and make the reconstruction more efficient.

In this paper we investigate the benefit of using detectors for faint light image acquisition. The particular questions we address is the minimum acquisition time (expressed as the minimum number of image frames or accumulated number of measurements) and the minimal absorbed number of photons per pixel element one needs to get a predefined contrast resolution in an image. 

The contrast between two signal-producing structures in an image is conventionally defined as the difference in signal (e.g. intensity or detected photon number) between the two objects divided by a reference signal, not seldom the sum of the two signals \cite{Hendrick}. However, in order to quantify the smallest contrast a detector can perceive, one also needs to consider the statistical variance (i.e. the noise) of the signals. The statistical variance is included in the contrast-to-noise ratio (CNR) that is defined as the the (absolute) difference in signal between the two objects divided by the standard deviation of this difference (assuming no statistical correlation between the signals) \cite{Timischl2015TheMicroscopy}. However, particularly in image recording of very weak signals, the signal's standard deviation is a function of the strength of the signal and the detector type. Thus, for faint light image acquisition, one needs a bit more elaborate analysis to determine the fundamental contrast resolution of a image recording system. This is the motivation behind this work.

In addition to determine the minimal number of frames or detected photon number needed to reach a certain contrast resolution, we also look at the error probability of correctly identifying one of many gray levels. In addition we ask if there is anything to gain by using photon number resolving pixel detectors instead of so-called click detectors that can only resolve between zero and more than zero absorbed photons. We will not look into the imaging system itself at all, but only look at the detector array (consisting of individual few-photon detectors) at the image plane. We will also exclude details about the image acquisition such as what wavelengths were used, what the acquisition time (or shutter speed) was, but have as our main parameter how many photons were detected. In this sense our study is generic, covering a wide range of different kinds of low intensity image acquisitions.

The paper is organized as follows. In Sec. \ref{sec:contrast-resolution} we define the resolution problem and derive the minimum number of absorbed photons to resolve a fixed number of gray levels in an image using a multiplexed, photon-number resolving (PNR) detector. This model can, depending on the chosen parameters, describe either a so-called click-detector (that is a detector that can only resolve between zero photons and one or more photons) or a linear PNR detector, and also cases in-between these extremes. 

In Sec \ref{sec:contrast-resolution-pnr} we analyze the performance of so-called intrinsic PNR detectors. Examples of such detectors are TES and superconducting tapered nanowire detectors (STaND) \cite{Zhu2019ResolvingDetector}. In Sec: \ref{sec:estimation-errors} we look at the probability of making level estimation errors due to the statistical nature of the detected photon number. In Sec. \ref{sec:optimal-level-spacing}  we derive how the gray levels to be distinguished should optimally be distributed to benefit maximally from the detectors' characteristics. Finally, in Sec. \ref{sec:conclusions}, we summarize our findings and draw conclusions.

\section{\label{sec:contrast-resolution}CONTRAST RESOLUTION OF FEW-PHOTON DETECTORS}
\begin{figure}[t]
    \centering
    \includegraphics[width=0.9\linewidth]{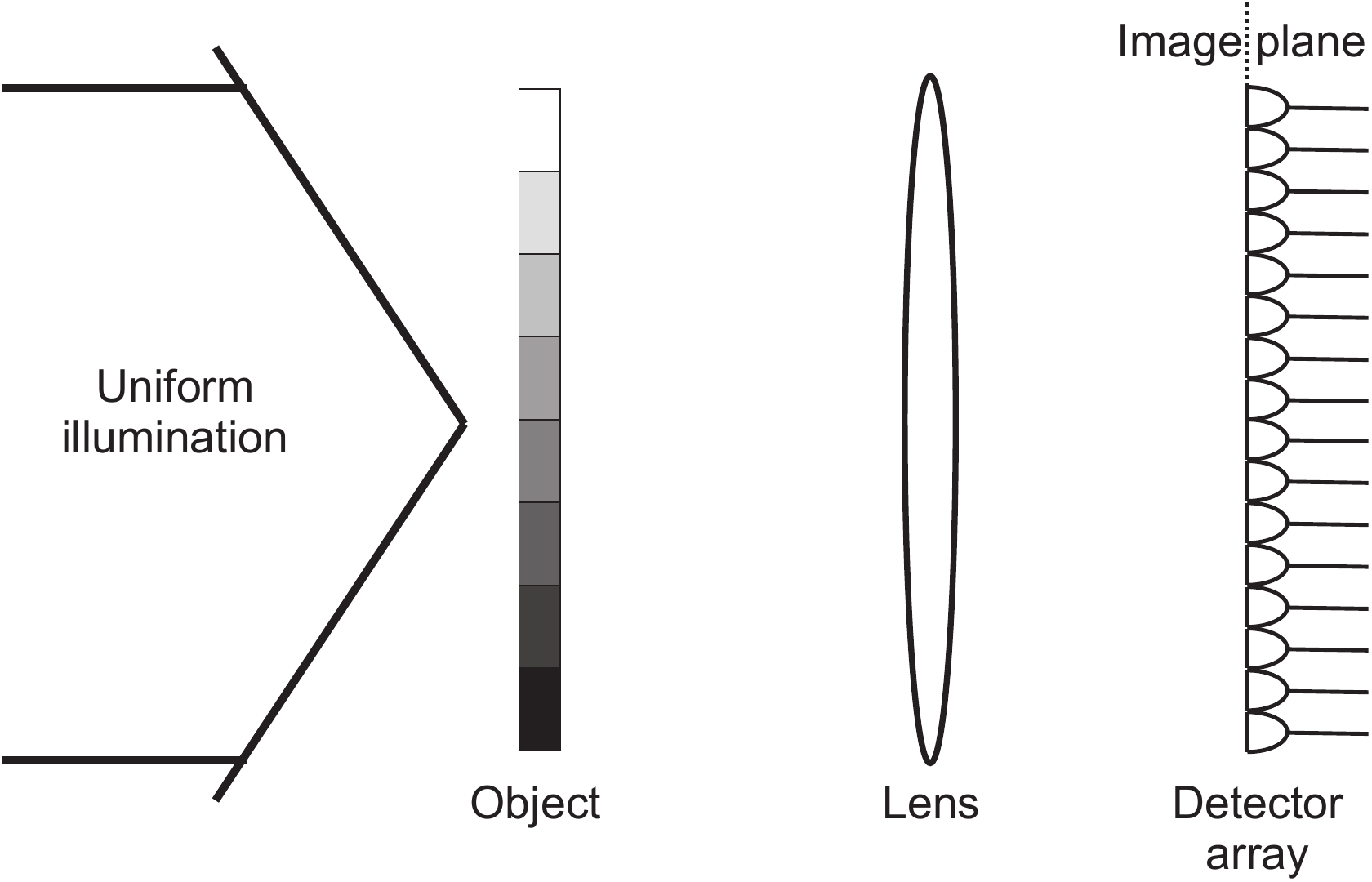}
    \caption{A schematic illustration of an illuminated object with different levels of transparency, image into a detector array.}
    \label{fig:setup}
\end{figure}

Suppose a semitransparent object is illuminated with a spatially uniform light source. The light that passes through the object is focused onto a detector array that, in an imaging scenario, would be called an array of pixels, see Fig. \ref{fig:setup}. In the image plane, the pixels are then illuminated by light of different intensities corresponding to the objects transparency at the imaged point. (A very similar imaging scenario would ensue for any illuminated, light-scattering object.)  To get a good gray-scale image, one needs to resolve between different light intensities, each intensity coding a certain transparency (or reflectance) of the object.

Suppose we want to resolve $L$ levels of intensity. (In many contemporary imaging applications the intensity is coded onto one byte, meaning that one discretize the intensity to $2^8 = 256$ different levels, translating to transparencies in our case.) We will initially assume that these intensities are equally spaced, meaning that we would like to resolve between the average transmitted photon numbers
\begin{equation}
    \mu(l) = \frac{\mu_0 l}{L - 1}, \quad l \in \{0, 1, 2, \dots, L - 1 \},
    \label{eq:levels}
\end{equation}
where $\mu_0$ is the mean illumination photon number per pixel and $l/(L-1)$ is the transmittance of level $l$.

To measure the transparency levels we consider each pixel in the array to be a multiplexed PNR detector with quantum efficiency $\eta$ consisting of $n$ elements, each with dark count probability $p_d$ per measurement. Since imaging necessitates detection of many modes, and typically photon number fluctuations in these modes during one acquisition frame are uncorrelated, the photon number fluctuations detected by any pixel can be modelled by a Poisson distribution. Under this assumption, the probability for a pixel to give $x \in \{0, 1, \dots, n\}$ clicks given the transparency level $l$ is (see appendix \ref{sec:derivation-of-the-poisson-pcd} for derivation)
\begin{equation}
    \Pr(x \mid l) = \binom{n}{x} (1 - p_d)^n e^{- \mu(l) \eta} \qty(\frac{e^{\mu(l) \eta / n}}{1 - p_d} - 1)^x,
    \label{eq:poisson-pnr-distribution}
\end{equation}
which is consistent with Ref. \cite{Zhu2019ResolvingDetector} when $p_d = 0$. This model is rather general in the sense that it can describe the case when each pixel is a single-photon detector by setting $n = 1$ and it can describe the case when each pixel is a linear PNR detector by letting $n \to \infty$ and setting $p_d = \nu/n$, where $\nu$ is the average number of dark counts per pixel and measurement.

To estimate the transparency levels we conduct $k$ measurements (i.e. we take $k$ image ``frames'') with the detector array and assume that the light intensity is constant from frame to frame. The result of the measurements is a sequence of data $\{X_i\}_{i = 1}^{k}$ for each pixel which can be used to compute an estimate of $l$. Using the maximum likelihood method to compute estimates yields that
\begin{equation}
    \hat{l} = - \frac{n (L - 1)}{\eta \mu_0} \ln(\frac{n - \ev{X}}{(1 - p_d) n}),
\end{equation}
where $\ev{X}$ is the average taken over the measurement data
\begin{equation}
    \ev{X} = \frac{1}{k} \sum_{i = 1}^k X_i.
    \label{eq:experimental-average}
\end{equation}
The estimator $\hat{l}$ is defined on the set $(0, \infty)$ while the level $l$ is defined on the set $\{0, 1, \dots, L - 1\}$, which means that we need to limit $\hat{l}$ to the smaller set. This can be done in the maximum likelihood sense by defining an discretized estimator to be
\begin{equation}
    \hat{l}' = \min \qty{ L - 1, \argmax_{l \in \floor{\hat{l}}, \ceil{\hat{l}}} \mathcal{L}(l)},
\end{equation}
where $\mathcal{L}$ is the likelihood function.

To ensure that the estimation is stable we require that the separation to the next level is $d$ times the standard deviation of the estimator, i.e. $1 \geq d \sqrt{\variance(\hat{l})}$. For large enough $d$ we get get that the levels are well-separated and the reconstructed image has the correct values with high probability. By using the Cramer-Rao bound we can get a lower bound on how many frames that are required to generate an image with separation $d$. We get that
\begin{equation}
    1 \geq d \sqrt{\variance(\hat{l})} \geq d \sqrt{\frac{1}{k I(l)}},
\end{equation}
where $I(l)$ is the Fisher information. Computing the Fisher information from equation \eqref{eq:poisson-pnr-distribution} from $k$ independent measurements gives that
\begin{equation}
    I(l) = \qty( \frac{\mu_0 \eta}{L - 1})^2 \frac{1}{n [(1 - p_d)^{-1} e^{\mu(l) \eta /n} - 1]}.
\end{equation}
Combining the equations gives a bound on the minimal number of measurements required for any $l$ to get $d$ standard deviations of separation under ideal conditions (see appendix \ref{sec:estimation} for details)
\begin{equation}
    k \geq \frac{n d^2 (L - 1)^2  \qty[(1 - p_d)^{-1} e^{\mu_0 \eta / n} - 1]}{(\mu_0 \eta)^2}.
    \label{eq:k-requirement}
\end{equation}

\begin{figure}[t]
    \centering
    \includegraphics[width=\linewidth]{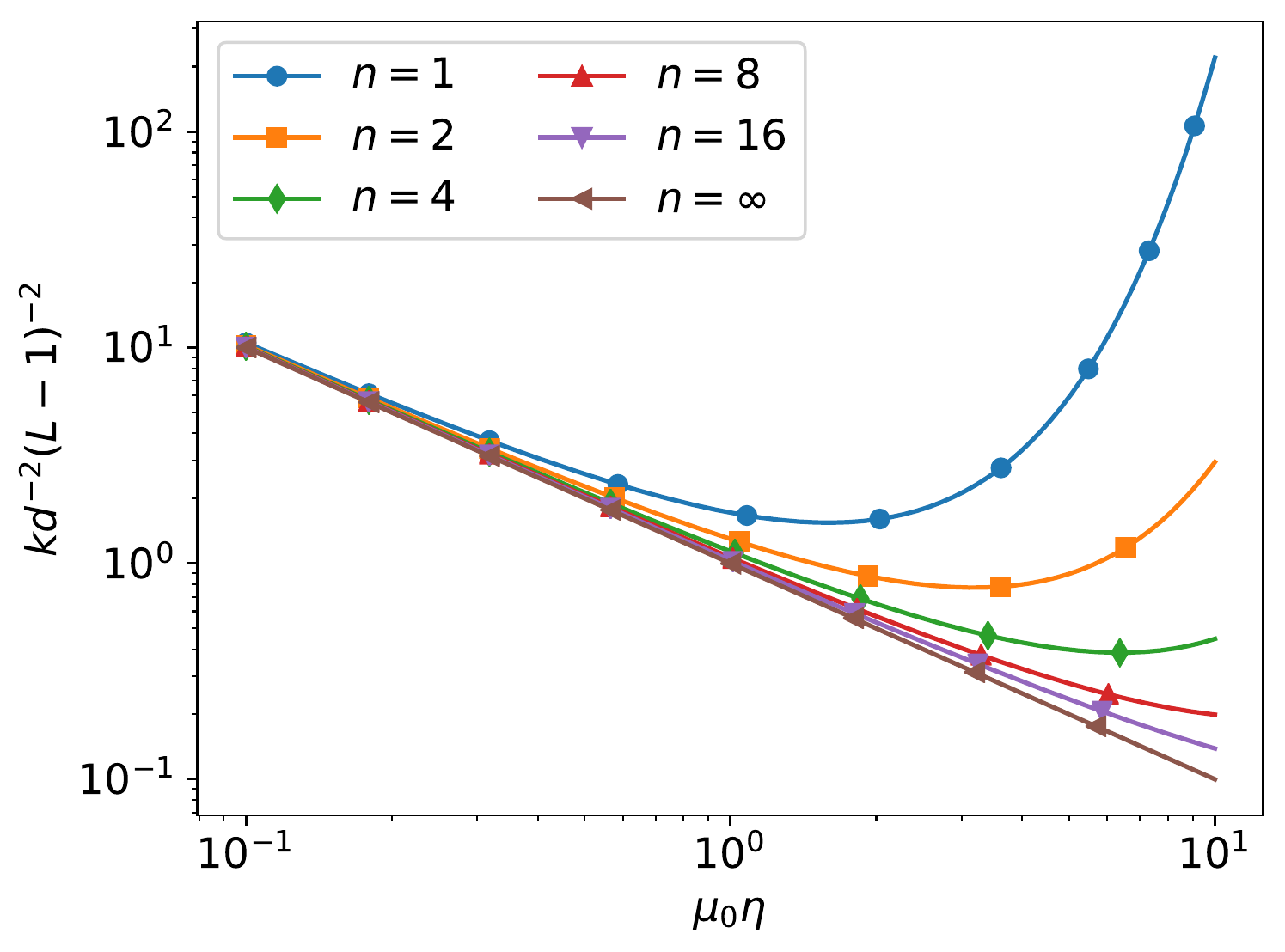}
    \caption{The number of required measurements $k$ plotted against the average mean photon number per pixel per frame $\mu_0 \eta$ for different number of elements $n$ in the multiplexed detectors and no dark counts. When $\mu_0 \eta \ll n$ there is no benefit to increase the number of detector elements, but when $\mu_0 \eta \gg n$ the number of measurements can be reduced significantly by increasing $n$. As expected, $k$ diverge when $\mu_0 \eta \to 0$ since the detector will always output $X = 0$ and amount of information gained per measurement is zero. Similarly $k$ diverges as $\mu_0 \eta \gg n$ since the detector always outputs $X = n$.}
    \label{fig:num-measurements}
\end{figure}

In Fig. \ref{fig:num-measurements} the number of required measurements $k$ is plotted against the detected mean photon number $\mu_0 \eta$ for different number of elements $n$. For low mean photon numbers $\mu_0 \eta \ll n$ all detectors with at least $n$ elements preform equally, while for high mean photon numbers $\mu_0 \eta \gg n$ a large number of elements is beneficial. This suggest that detectors with capability to resolve more than one photon are beneficial for estimating high mean photon numbers.

\begin{figure}[t]
    \centering
    \includegraphics[width=\linewidth]{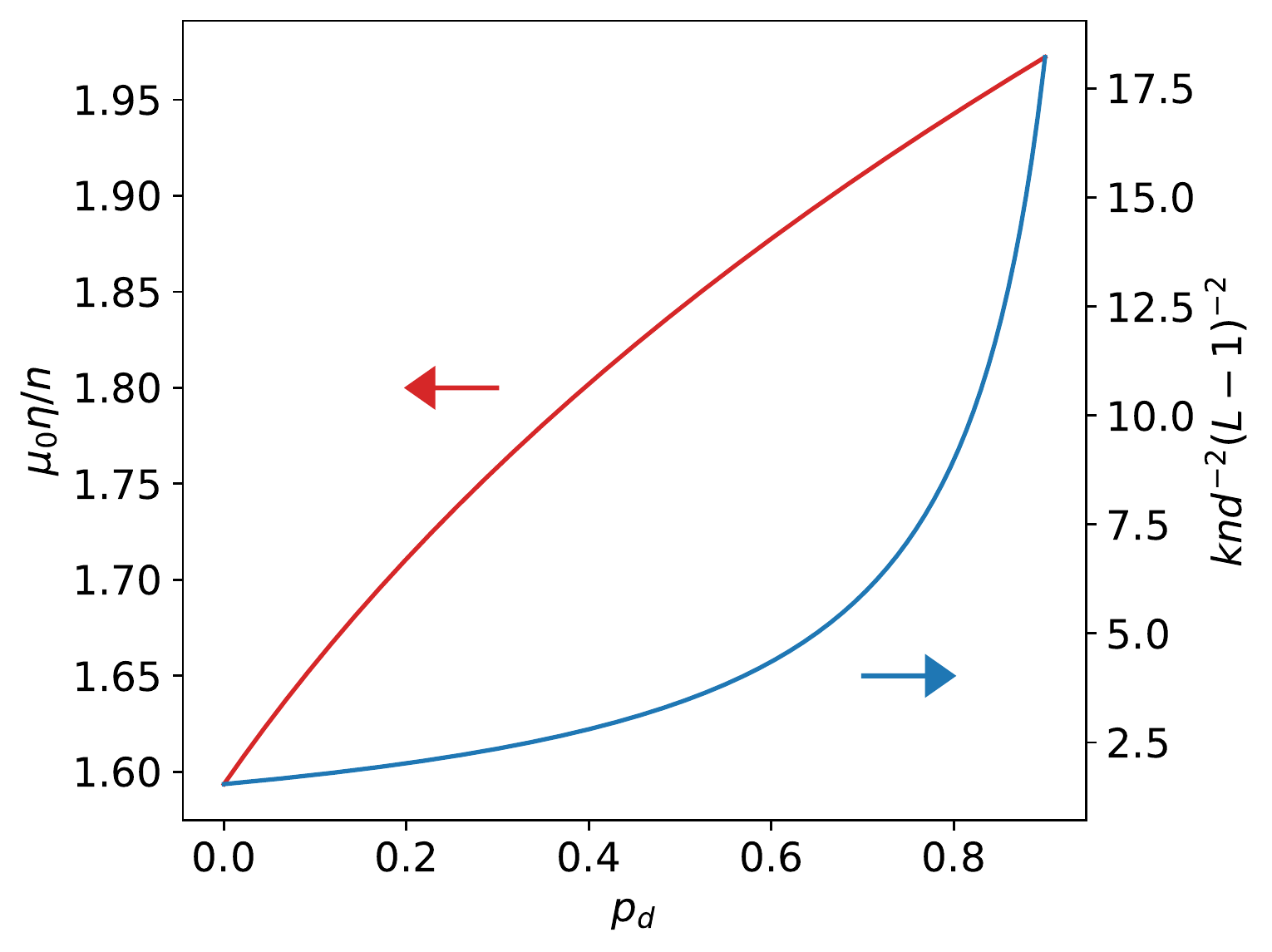}
    \caption{The optimal illumination as a function of the dark count probability. The arrows mark which axis corresponds to which curve. The upper red curve shows the mean photon number that minimizes the number of measurements required. The lower blue curve shows the number of measurements required to achieve $d$ standard deviations of separation.}
    \label{fig:optimal-illumination}
\end{figure}

As seen in Fig. \ref{fig:num-measurements} there exists a mean photon number that minimizes the number of measurements required. This minimum is achieved when the mean photon number is selected as
\begin{equation}
    \mu_0 \eta = n \qty[ 2 + W_0\qty( -2 (1 - p_d) e^{-2})],
    \label{eq:minima-condition}
\end{equation}
where $W_0$ is the Lambert W function. When $p_d = 0$ the optimal mean photon number is given by $\mu_0 \eta = 1.5936n$, while when $p_d \to 1$ the minimum is given by $\mu_0 \eta \to 2n$, see Fig. \ref{fig:optimal-illumination}. 

\section{\label{sec:contrast-resolution-pnr}CONTRAST RESOLUTION OF INTRINSIC PNR DETECTORS}

\begin{figure}
    \centering
    \includegraphics[width=\linewidth]{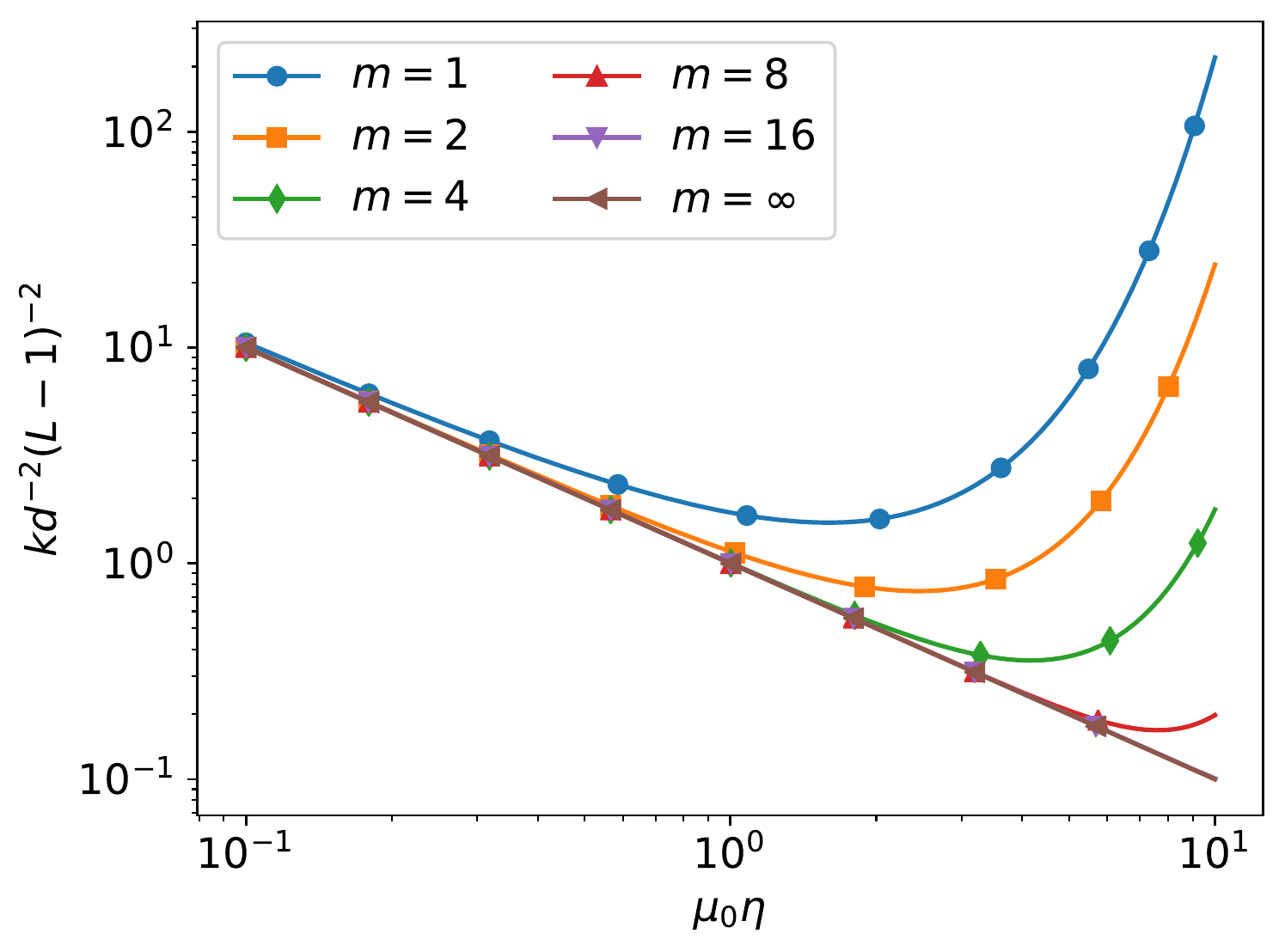}
    \caption{The number of required measurements $k$ plotted against the effective mean photon number $\mu_0 \eta$ for an intrinsic PNR detector ($n=\infty$) with a hard cut-off at $m$ photons. The behavior is similar to the case when multiplexed detectors are used. However, the gain of increasing the number of resolvable photons is smaller than in the multiplexed case.}
    \label{fig:hard-cut-off}
\end{figure}

Intrinsic PNR detectors such as TES or STaND can be seen as multiplexed detectors with a large number of elements. To a good approximation we can consider them to have an infinite number of elements, which together with equation \eqref{eq:poisson-pnr-distribution} gives that the probability to get $x$ as output given the transparency level $l$ is a Poisson distribution given by
\begin{equation}
    \Pr(x \mid l) = e^{- \mu(l) \eta - \nu} \frac{\qty[\mu(l) \eta + \nu]^x}{x!},
    \label{eq:poisson-output}
\end{equation}
where $\nu$ is the mean number of dark counts given by the relation $\nu = p_d  n$. Similarly we can show that the required number of measurements to achieve $d$ standard deviations of separation between levels is given by
\begin{equation}
    k \geq \frac{d^2 (L - 1)^2}{\mu_0 \eta} \qty( 1 + \frac{\nu}{\mu_0 \eta}).
\end{equation}
Unlike the case for finite $n$, the number of measurements $k$ does not have a minimum for a finite mean photon number and the optimal strategy according to the model is to use maximal illumination. However, in reality these PNR detectors are limited to how many photons they are capable of resolving due to overlapping output signals or non-linear output signals \cite{Zhu2019ResolvingDetector, Nicolich2019UniversalDetectors, Humphreys_2015}, which implies that there is an upper limit for when increased illumination is beneficial. To capture this behavior we need to consider a refined model where these effects are included.

A simple but somewhat crude model that describes this limitation can be achieved by introducing a hard cut-off on how many photons the detector is capable of measuring. This is done by assuming that inputs with up to $m$ photons are resolvable, while photon numbers higher than $m$ produce outputs indistinguishable from a $m$ photon event. The qualitative behavior of the intrinsic PNR detector can then be investigated by setting an upper photon number limit for which the detector is not capable of resolving for some reason. 

Using the hard cut-off model results in an output distribution which is given by equation \eqref{eq:poisson-output} when $x < m$,
\begin{equation}
    \Pr(x \mid l) = \sum_{y \geq m} e^{- \mu(l) \eta - \nu} \frac{[\mu(l) \eta + \nu]^y}{y!},
\end{equation}
for $x = m$ and zero for $x > m$. Hence, the detector is acting as a linear PNR detector up to $m$ photons and is then not capable of distinguishing between higher photon numbers due to the cut-off. It is therefore expected that the detector preforms identically to a linear PNR detector for small enough mean photon numbers.

Given $k$ data points from the intrinsic cut-off detector we can estimate the level $l$ using an estimator $\hat{l}$. Requiring that the estimator has a standard deviation $d$ times less than the separation to the next level gives a condition on the minimal number of measurements $k$, see Fig. \ref{fig:hard-cut-off}. The qualitative result is similar to the multiplexed case, but the gain of increasing $m$ is not as significant as increasing the number of detector elements in the multiplexed case. However, the acquisition time can be significantly reduced by using the extra information gained by measuring photon numbers.

\section{\label{sec:estimation-errors}ESTIMATION ERRORS}
Above we have used the standard deviation as a parameter to determine how well a detector is able to distinguish between different gray-levels. Having a larger number of standard deviations between neighbouring levels reduces the risk for incorrect classification, but we have so far not specified how the number of standard deviations translates to the error probability. Here, we want to quantify the error in the case when detectors described by equation \eqref{eq:poisson-pnr-distribution} is used.

The probability for a correct classification is given by the probability that the discretized estimator $\hat{l}'$ is equal to the correct level $l$. To determine this probability we use that the discrete estimator $\hat{l}'$ is a function of the experimental mean $\ev{X}$, which gives that the probability for correct classification is
\begin{equation}
    \Pr(\hat{l}' = a \mid l = a) = \Pr(\ev{X} \in S(a) \mid l = a),
\end{equation}
where $S(a)$ is the set of values for which $\ev{X}$ produces the discretized estimator $\hat{l}' = a$. This set is given by all $\ev{X}$ that makes the likelihood function maximal for the level $l = a$ and it can be determined using the conditions
\begin{align}
    \mathcal{L}(a) &\geq \mathcal{L}(a + 1),\\
    \mathcal{L}(a) &\geq \mathcal{L}(a - 1).
\end{align}
These conditions gives that for $0 < a < L - 1$ the set is given by
\begin{equation}
    S(a) = \qty(\frac{\mu_0 \eta (L - 1)^{-1}}{\ln[g(a) / g(a - 1)]}, \frac{\mu_0 \eta (L - 1)^{-1}}{\ln[g(a + 1) / g(a)]}),
\end{equation}
where the function 
\begin{equation}
    g(a) = e^{\mu(a) \eta / n} - (1 - p_d).
\end{equation}
When $a = 0$ the the lower limit for $S(0)$ is given by $0$ which the minimal value for $\ev{X}$. Hence 
\begin{equation}
    S(0) = \qty(0, \frac{\mu_0 \eta (L - 1)^{-1}}{\ln[g(1) / p_d]}).
\end{equation}
Similarly, when $a = L - 1$ the upper limit for $S(L - 1)$ is given by $n$ which is the maximal value for $\ev{X}$. Hence,
\begin{equation}
    S(L - 1) = \qty(\frac{\mu_0 \eta (L - 1)^{-1}}{\ln[g(L - 1) / g(L - 2)]}, n).
\end{equation}

The probability that the experimental average is in the set $S(a)$ can be estimated using the central limit theorem, which states that the experimental average converges in distribution to a normal distribution as the number of acquisitions grows, i.e.
\begin{equation}
    \ev{X} \rightharpoonup \mathcal{N}\qty(\mathbb{E}[X], \frac{\variance(X)}{k}), \text{ as } k \to \infty,
\end{equation}
where $\mathbb{E}[X]$ and $\variance(X)$ is the expected value and variance of $X$, respectively. For sufficiently large $k$ it holds approximately that
\begin{equation}
    \Pr(\ev{X} \in S(a) \mid l = a) \approx \Pr(\nu \in S(a) \mid l = a),
\end{equation}
where $\nu \sim \mathcal{N}\qty(\mathbb{E}[X], \variance(X) / k)$ is a normally distributed variable with the same average as the random variable $X$ and variance equal to the variance of $X$ divided by the number of acquisitions.

\begin{figure}[t]
    \centering
    \includegraphics[width=\linewidth]{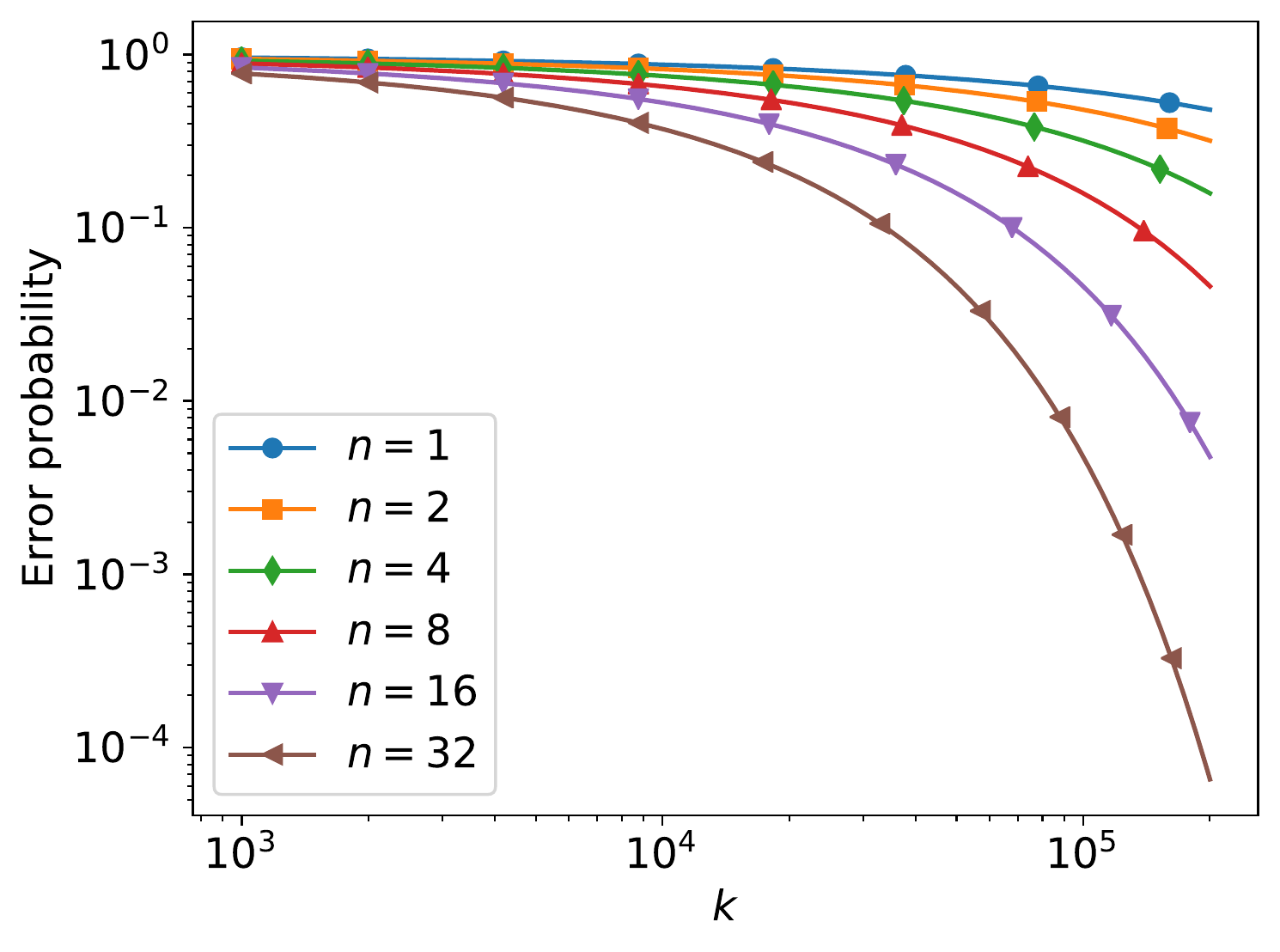}
    \caption{The probability for an incorrect classification of gray level for the most difficult level to resolve (level $L -1$) as a function of the number of acquisitions, when the mean photon number fulfills the condition in equation \eqref{eq:minima-condition}.}
    \label{fig:error-vs-k}
\end{figure}

\begin{figure}[t]
    \centering
    \includegraphics[width=\linewidth]{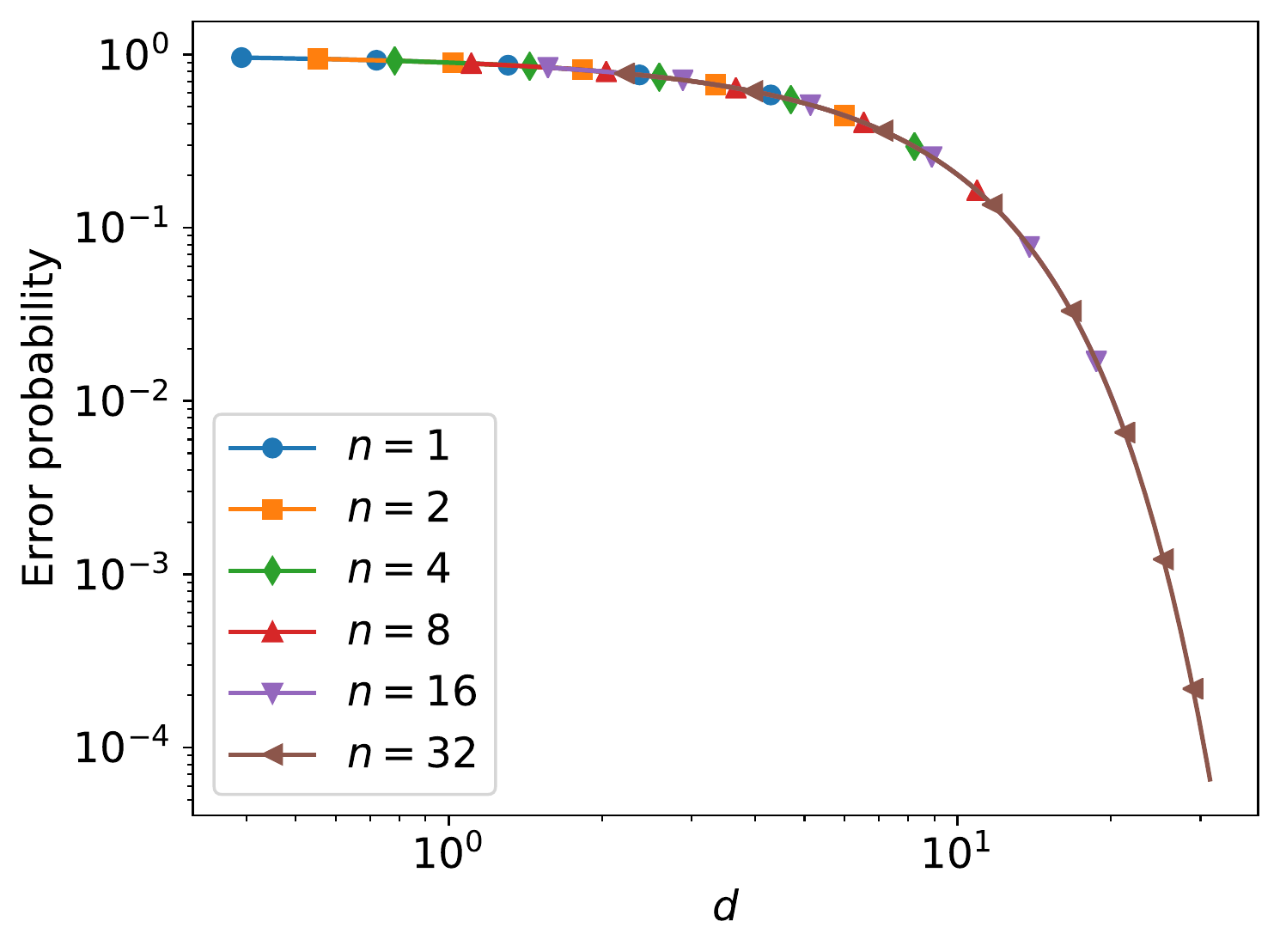}
    \caption{The probability for an incorrect classification for the most difficult level to resolve (level $L -1$) as a function of the number of standard deviations between two levels when the mean photon number fulfills the condition in equation \eqref{eq:minima-condition}. For this particular choice of mean photon number the error is independent of the number of detector elements $n$.}
    \label{fig:error-vs-d}
\end{figure}

In Fig. \ref{fig:error-vs-k} the error probability as a function of the number of acquisitions $k$ for the most difficult levels to resolve (levels $L-2$ and $L-1$) is presented when the mean photon number has been chosen to minimize $k$ [see equation \eqref{eq:minima-condition}]. The error probability decreases with increased $k$ and increased number of detector elements $n$. In Fig. \ref{fig:error-vs-d} the error probability as a function of the number of standard deviations $d$ between two levels is presented when the mean photon number per frame $\mu_0 \eta$ has been chosen to minimize $k$. Curves are overlapping which shows that the error probability is uniquely determined by $d$. At this optimum $\mu_0 \eta$ the total exposure $k \mu_0 \eta$ is equal to $- [d(L-1)]^2 / W_0(-2 (1 - p_d) e^{-2})$ for any $n$.

\section{\label{sec:optimal-level-spacing}OPTIMAL GRAY LEVEL SPACING}

\begin{figure}[t]
    \centering
    \includegraphics[width=\linewidth]{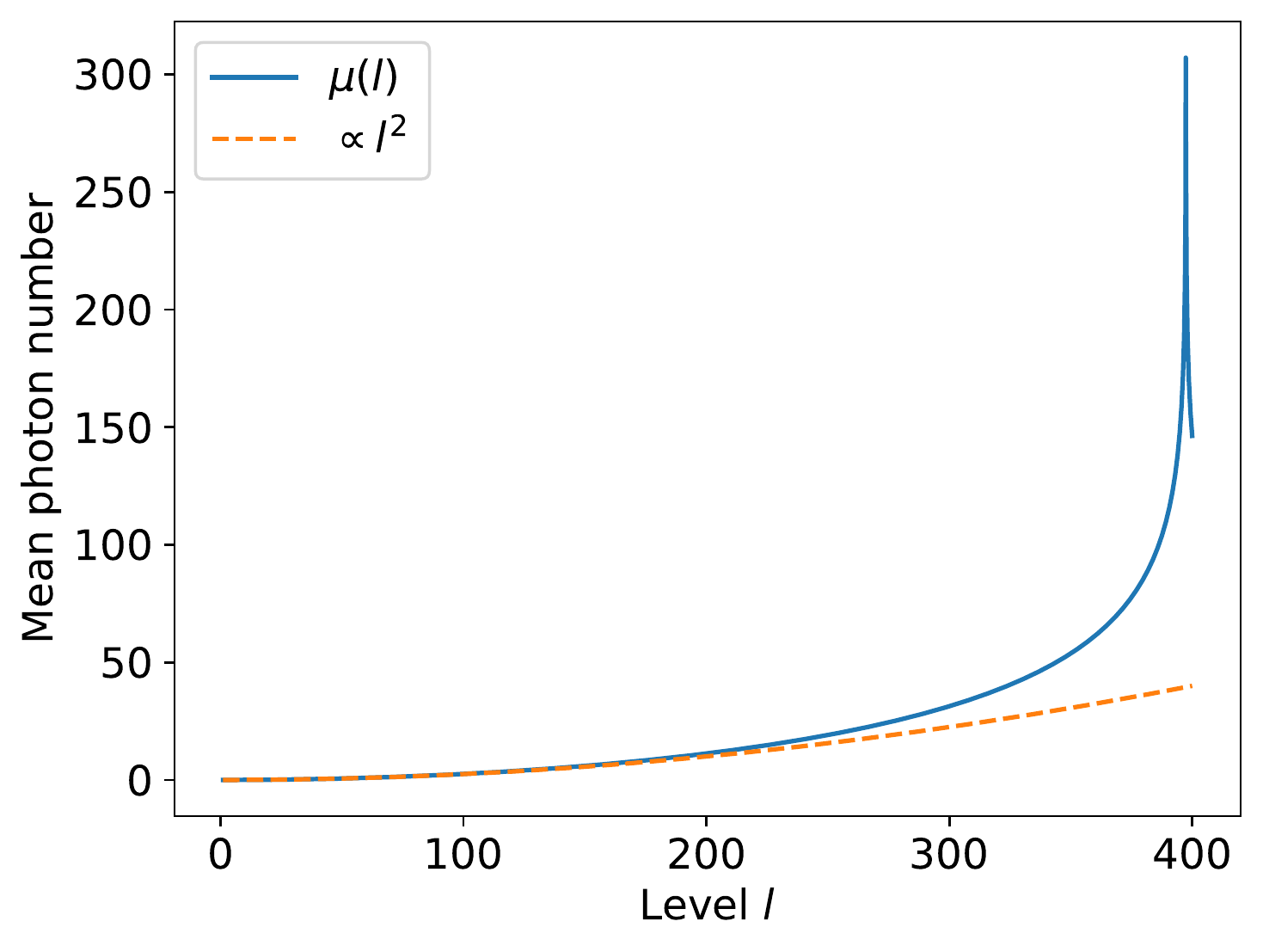}
    \caption{Optimal level spacing when $n = 16$, $d = 1$, $k = \SI{1e3}{}$ and no dark counts. The mean photon number diverges above the maximal level $l_{\text{max}} = 397$ and the solution is therefore not valid above this level. For the levels with sufficiently small level index $l$, the function $\mu(l)$ can be approximated by $(l d)^2 / 4 k$.}
    \label{fig:optimal-non-linear-spacing}
\end{figure}

\begin{figure}[t]
    \centering
    \includegraphics[width=\linewidth]{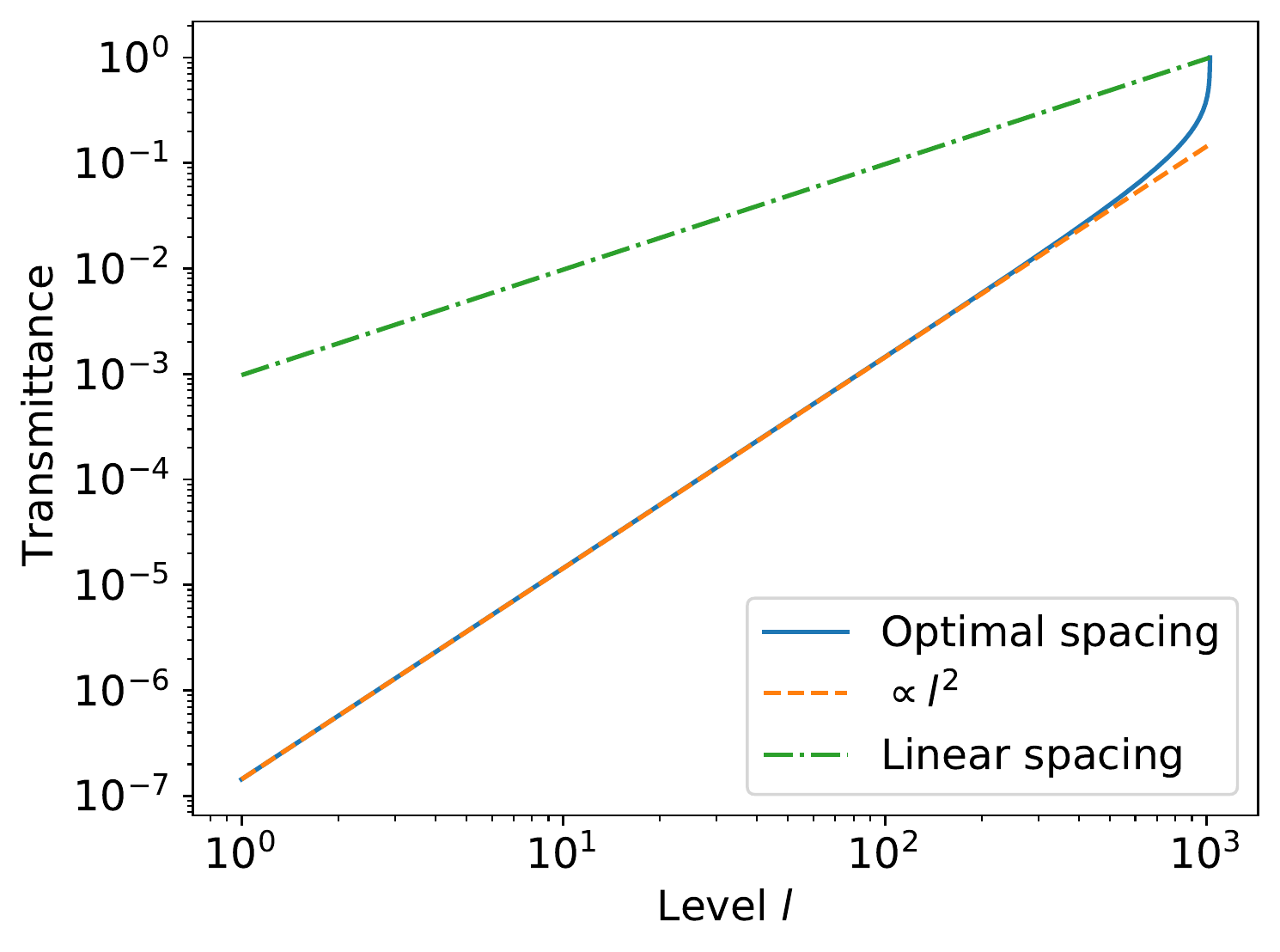}
    \caption{Transmittance as a function of the level when the maximal level is $1024$ for optimal spacing, and linear spacing. The optimal spacing grows quadratically for sufficiently small levels.}
    \label{fig:optimal-vs-linear-spacing}
\end{figure}

In previous sections we have considered gray levels equally spaced, and with the assumption that the level $L - 1$  corresponds to unit transmittance. When setting a requirement on how many measurements (or frames) $k$ are needed to achieve $d$ standard deviations separation of neighbouring levels this is not the best scenario. With equal level spacing, darker levels are easier to resolve than lighter levels, and it should therefore be possible to improve the contrast resolution by allowing non-linear level spacing.

The optimal level spacing occurs when every level is limiting for the number of acquisitions $k$ given some requirement $d$ on the number of standard deviations separating two neighbouring levels. From the condition in equation \eqref{eq:k-requirement} we get a differential equation
\begin{equation}
    \qty(\pdv{\mu(l)}{l})^2 \frac{\eta^2}{n \qty[(1 - p_d)^{-1} e^{\mu(l) \eta / n} - 1]} = \frac{d^2}{k},
\end{equation}
which holds for all levels $l \in \{0, 1, \ldots, L - 1 \}$. Let us assume that first level is at zero intensity, i.e. $\mu(l = 0) = 0$, then the solution to the differential equation is given by
\begin{equation}
\begin{split}
    \mu(l) = & \frac{n}{\eta} \ln\qty[1 + \tan^2\qty(\frac{l d}{2 \sqrt{kn}} + \arctan(\sqrt{\frac{p_d}{1 - p_d}}))]\\
    & + \frac{n}{\eta} \ln(1 - p_d),
    \label{eq:non-linear-optimal-spacing}
\end{split}
\end{equation}
given that the condition $d^2/(4 \eta)>1$ is met, which for most realistic situations should be the case. If not, some levels with small index $l$ will have $k \mu(l) \leq 1$, and since the measurement outcome is always an integer, and for these levels the most likely outcome are zero or one, the levels cannot be distinguished.

Let the parameters $\eta, n, d, k$ and $p_d$ be fixed, then there is a finite number of resolvable gray levels. Increasing the number of levels further would violate the condition that there should be $d$ standard deviations separation between levels. The maximal level is found by investigating where the solution $\mu(l)$ stops being valid, which occurs when the solution diverges. The maximal level is therefore given by
\begin{equation}
\begin{split}
    l_{\text{max}} &= \max \qty{l \in \mathbb{N} \mid \abs{\mu(x)} < \infty, \; \forall x \in [0, l]}\\
    &= \bigg\lfloor\frac{2 \sqrt{kn}}{d} \qty[\frac{\pi}{2} - \arctan\qty(\sqrt{\frac{p_d}{1 - p_d}})]\bigg\rfloor.
\end{split}
\end{equation}
This equation shows the advantage of PNR detectors, since the maximum number of distinguishable gray levels increase as $n^{1/2}$, that is, essentially with the square root of the detector's photon number resolving capability. We also see that the detector's quantum efficiency does not limit its resolution capability, whereas it's dark count probability does.

In Fig. \ref{fig:optimal-non-linear-spacing} the mean photon numbers $\mu(l)$ is plotted as a function of the level $l$. As predicted the mean photon number diverges in the interval $l \in (l_{\text{max}}, l_{\text{max}} + 1]$ and the solution is no longer valid for levels larger than $l_{\text{max}}$. 

Taking the ratio $\mu(l)/\mu(l_{\text{max}}-1)$ we get the transmittance of level $l$. This is plotted in Fig. \ref{fig:optimal-vs-linear-spacing} assuming that $p_d=0$. In the plot we have chosen $\sqrt{k n}/d \approx 326$ to get $\mu(l_{\text{max}}-1)=1024$. For sufficiently small transmittance (or level indices $l$) the transmittance grows quadratically with the level index. As a reference we have plotted 1024 equally spaced transmittancies, with the highest being unity. Choosing the levels equally spaced results in a poorer performance, because as explained in Sec. \ref{sec:contrast-resolution}, the variance of the measured signal is largest for the highest indices $l$ and therefore one should try to maximize the transmittance difference, translating to maximizing the mean average transmitted photon number difference, at the highest level indices. The discrepancy between the two spacing models can clearly be seen by comparing the derivatives of the two functions at $l_{\text{max}}$.

\section{\label{sec:conclusions}SUMMARY}

We have shown that in order to resolve contrast in a monochrome image, and thereby extract the information contained therein, it could be advantageous to use PNR detectors as detector elements in the image detector array. Provided that the object is illuminated so that the average number of detected photons per pixel (detector) per image frame is above unity, and within the photon number resolving range of the PNR detectors,  PNR detectors will provide a better contrast resolution than even an ideal, click-detector-based image acquisition system. The relative difference in performance between the two detector types diverge exponentially with increasing number of detected photons per pixel per frame. 

We have shown that the performance in terms of needed measurements (i.e. number of accumulated image frames) for a given acceptable error probability of a PNR detector decreases linearly with the average number of detected photons per pixel (detector) per image frame, given that the number is within the range resolvable by the detector. In contradistinction, a click-detector-based system has an optimal detected photon number per pixel (detector) per image frame. Below or above this optimum less information is extracted for a given number of measurements.

We have also quantified the probability of incorrect classification of the gray levels. It was shown that for a PNR detector this error probability is independent of the range of photons the detector can distinguish between, provided that the detected photons per pixel (detector) per image ``frame'' is chosen optimally. The advantage of using a detector with larger PNR capability is that the number of needed measurements (or frames) decreases with increasing $n$. Since the statistical variation at the pixel level between frames is assumed to be uncorrelated, the accumulated needed transmitted photon number is independent of $n$, implying that under the assumptions just stated, the needed number of measurements $k$ scales as $\propto 1/n$.

Finally we have shown, not surprisingly, that to resolve maximally many gray levels with a given error probability (i.e. a certain $d$), the levels should not be equidistant in transmittance but follow a non-linear level spacing. For the darkest shades, the transmittance should increase approximately quadratically with the level number $l$. This is in line with the human vision, where the eye is significantly better at distinguishing closely spaced levels of dark gray (faint light) than equally spaced levels of light gray (intense light) \cite{kimpe2007}.

\begin{acknowledgments}
This work was supported by Knut and Alice Wallenberg Foundation through the grant Quantum Sensing and by the Swedish Research Council (VR) through grant 621-2014-5410.
\end{acknowledgments}

\appendix

\section{\label{sec:derivation-of-the-poisson-pcd}DERIVATION THE POISSON PHOTON COUNTING DISTRIBUTION}
Consider a multiplexed PNR detector with $n$ elements with quantum efficiency $\eta$ and dark count probability $p_d$. In a previous work \cite{Jonsson2020Photon-countingDetectors} we showed that the probability to get $x \in \mathbb{N}$ clicks when a number state $\ket{m}$ is incident is given by
\begin{equation}
\begin{split}
    \Pr(x \mid m) &= \frac{1}{n^m} \binom{n}{x} \sum_{l = 0}^x (-1)^l (1 - p_d)^{n - x + l} \binom{x}{l} \times\\
    &\times [n - (n - x + l)\eta]^m. 
\end{split}
\end{equation}
Using this distribution we want derive the resulting derivation for when a coherent state with mean photon number $\mu$ is incident. Using that the coherent state has Poisson statistics gives that the probability to get $x \in \mathbb{N}$ is
\begin{equation}
\begin{split}
    \Pr(x \mid \mu) &= \sum_{m \in \mathbb{N}} \frac{e^{-\mu} \mu^m}{m!} \Pr(x \mid m)\\
    &= e^{-\mu} \binom{n}{x} \sum_{l = 0}^x (-1)^l (1 - p_d)^{n - x + l} \times\\
    &\times \binom{x}{l} \sum_{m \in \mathbb{N}} \frac{1}{m!} \bigg( \frac{\mu [n - (n - x + l)]}{n} \bigg)^m.
\end{split}
\end{equation}
Multiplying the inner sum with $\exp{-\mu [n - (n - x + l)] / n}$ and using that the Poisson distribution sums to one gives that
\begin{equation}
\begin{split}
    \Pr(x \mid \mu) &= \binom{n}{x} (1 - p_d)^{n - x} e^{-\mu \eta (1 - x / n)} \times\\
    &\times \sum_{l = 0}^x \binom{x}{l} \qty[-(1 - p_d) e^{-\mu \eta / n}]^l 1^{x - l}\\
    &= \binom{n}{x} (1 - p_d)^n e^{- \mu \eta} \qty( \frac{e^{\mu \eta / n}}{1 - p_d} - 1)^x,
    \label{eq:derived-poisson-pcd}
\end{split}
\end{equation}
where the binomial theorem was used to compute the sum in the last equality.

The resulting probability distribution has the expected value 
\begin{equation}
    \mathbb{E}[x] = n \qty[1 - (1 - p_d) e^{-\mu \eta / n}]
\end{equation}
and the variance
\begin{equation}
\begin{split}
    \variance(x) &= \mathbb{E}[x^2] - \mathbb{E}[x]^2\\
    &= n \big[ 1 - (1 - p_d) e^{-\mu \eta / n} \big] (1 - p_d) e^{-\mu \eta / n}.
    \label{eq:variance}
\end{split}
\end{equation}

The Fisher information for the probability distribution given in equation \eqref{eq:derived-poisson-pcd} is given by
\begin{equation}
\begin{split}
    I(\mu) &= \mathbb{E}\qty[\qty(\pdv{\ln \Pr(x \mid \mu)}{\mu})^2]\\
    &= \variance \qty(\pdv{\ln \Pr(x \mid \mu)}{\mu}),
\end{split}
\end{equation}
where we have used that the score has zero expected value. The score is given by
\begin{equation}
    \pdv{\ln \Pr(x \mid \mu)}{\mu} = \frac{\eta x}{n \qty(1 - (1 - p_d) e^{-\mu \eta / n})} - \eta,
    \label{eq:score}
\end{equation}
which together with equation \eqref{eq:variance} gives the Fisher information
\begin{equation}
\begin{split}
    I(\mu) &= \frac{\eta^2 \variance(x)}{n^2 \qty(1 - (1 - p_d) e^{-\mu \eta / n})^2}\\
    &= \frac{\eta^2}{n \qty[(1 - p_d)^{-1} e^{\mu \eta / n} - 1]}.
    \label{eq:fisher-information-mu}
\end{split}
\end{equation}

\section{\label{sec:estimation}ESTIMATORS AND VARIANCE}
Here we derive an expression for the maximal likelihood estimator $\hat{l}$ from a set of data $\{ X_i \}_{i = 1}^k$ given that the illumination is given by equation \eqref{eq:levels} and the output distribution is given by equation \eqref{eq:poisson-pnr-distribution}. Using equation \eqref{eq:score} we get that the estimator is given by the condition
\begin{equation}
\begin{split}
    0 &= \sum_{i = 1}^k \pdv{\ln{\Pr(X_i \mid \mu(\hat{l}))}}{\mu(\hat{l})} \pdv{\mu(\hat{l})}{\hat{l}}\\
    &= \frac{\mu_0}{L - 1} \qty[ \frac{\eta \sum_{i = 1}^k X_i}{n \qty(1 - (1 - p_d) e^{-\mu(\hat{l}) \eta / n})} - k \eta ].
\end{split}
\end{equation}
Inverting the expression and introducing the average over the experimental data $\ev{X}$ as equation \eqref{eq:experimental-average} gives that
\begin{equation}
    \hat{l} = - \frac{n (L - 1)}{\eta \mu_0} \ln(\frac{n - \ev{X}}{(1 - p_d) n}).
\end{equation}

The maximal likelihood estimator $\hat{l}$ is asymptotically unbiased \cite{amari2016information} and the Cramer-Rao bound can therefore be used to give a lower bound on the variance
\begin{equation}
\begin{split}
    \variance(\hat{l}) &\geq \frac{1}{k I(l)} = \frac{1}{k I\qty(\mu(l))} \qty(\pdv{\mu(l)}{l})^{-2}\\
    &= \frac{n (L - 1)^2 \qty[(1 - p_d)^{-1} e^{\mu(l) \eta / n} - 1]}{k (\mu_0 \eta)^2},
\end{split}
\end{equation}
where we have used the variable substitution role for Fisher information and equation \eqref{eq:fisher-information-mu}.

Requiring $d$ standard deviations of separation between peaks implies that
\begin{equation}
    k \geq \frac{n d^2 (L - 1)^2  \qty[(1 - p_d)^{-1} e^{\mu(l) \eta / n} - 1]}{(\mu_0 \eta)^2},
\end{equation}
holds for every $l \in \{0, 1, \dots, L - 1\}$. If the bound holds for $l = L - 1$ then it automatically holds for any other $l$ an we can therefore simplify the expression by setting $\mu(l) = \mu_0$, which gives equation \eqref{eq:k-requirement}.

To minimize $k$ with respect to $\mu_0 \eta$ is equivalent to minimize the function 
\begin{equation}
    f(\xi) = \frac{\alpha e^{\xi / n} - 1}{\xi^2},
\end{equation}
where we have set $\xi = \mu_0 \eta$, $\alpha = (1 - p_d)^{-1}$. The minimum is occurs when $\dv*{f}{\xi} = 0$ which corresponds to 
\begin{equation}
    \alpha e^{\xi/n} - 2 n \alpha e^{\xi/n} + 2 n = 0,
\end{equation}
which under the constraint $\xi \in [0, \infty)$ has the solution
\begin{equation}
    \xi = n \qty[ 2 + W_0\qty(-\frac{2}{\alpha} e^{-2}) ],
\end{equation}
where $W_0$ is the Lambert W function. Substituting back the variables gives the condition for the minimum in equation \eqref{eq:minima-condition}.
\end{document}